\documentclass{emulateapj}

\newcommand{\beq}{\begin{equation}}
\newcommand{\beqa}{\begin{eqnarray}}
\newcommand{\eeq}{\end{equation}}
\newcommand{\eeqa}{\end{eqnarray}}
\def\vperp{{\bf v}_\perp}
\def\vpar{{v}_\parallel}
\def\aperp{{\bf \dot{v}}_\perp}
\def\apar{{\dot v}_\parallel}
\def\HD{HD 209458b~}

\begin{document}

\title{A Dynamical Method for Measuring Masses of Stars with Transiting
Planets}

\author{Abraham Loeb}

\affil{Astronomy Department, Harvard University, 60 Garden Street,
Cambridge, MA 02138; aloeb@cfa.harvard.edu}

\begin{abstract}

As a planet transits the face of a star, it accelerates along the
line-of-sight.  The changing delay in the propagation of photons produces
an apparent deceleration of the planet across the sky throughout the
transit.  This persistent transverse deceleration breaks the time-reversal
symmetry in the transit lightcurve of a spherical planet in a circular
orbit around a perfectly symmetric star.  For ``hot Jupiter'' systems,
ingress advances at a higher rate than egress by a fraction $\sim
10^{-4}$--$10^{-3}$.  Forthcoming space telescopes such as {\it Kepler} or
{\it COROT} will reach the sensitivity required to detect this asymmetry.
The scaling of the fractional asymmetry with stellar mass $M_\star$ and
planetary orbital radius $a$ as $\propto M_\star/a^2$ is different from
that of the orbital period as $\propto (M_\star/a^3)^{-1/2}$. Therefore,
this effect constitutes a new method for a purely dynamical determination
of the mass of the star, which is currently inferred indirectly with
theoretical uncertainties based on spectral modeling. Radial velocity data
for the reflex motion of the star can then be used to determine the
planet's mass.  Although orbital eccentricity could introduce a larger
asymmetry than the light propagation delay, the eccentricity is expected to
decay by tidal dissipation to negligible values for a close-in planet with
no perturbing third body. Future detection of the eclipse of a planet's
emission by its star could be used to measure the light propagation delay
across the orbital diameter, $46.7 (a/7\times 10^{11}~{\rm cm})$ seconds,
and also determine the stellar mass from the orbital period.

\end{abstract}


\section{Introduction}

The population of known extrasolar planets which transit the face of their
parent stars has been growing steadily in recent years. It currently
includes \HD \citep{Cha}, OGLE-TR-56b \citep{Tor}, OGLE-TR-113b
\citep{Kon}, OGLE-TR-132b \citep{Mou}, TrES-1 \citep{Alo}, OGLE-TR-111b
\citep{Pon} and OGLE-TR10b \citep{Kon2}.  For \HD, the transit lightcurve
was observed with the {\it Hubble Space Telescope (HST)} to be symmetric
around its centroid to an exquisite photometric precision of $\sim 10^{-4}$
magnitude per data point for a few hundred data points tracing both ingress
and egress \citep{Bro}.

In this {\it Letter} we show that transit lightcurves must have a
time-reversal asymmetry even if the star and the planet are perfectly
symmetric and the orbit is circular.  The asymmetry originates from the
persistent acceleration of the planet towards the star during the
transit. For a circular orbit, the planet moves towards the observer at the
beginning of the transit (ingress) and away from the observer at its end
(egress). This net acceleration introduces a change in the relative rate by
which ingress and egress advance in the observer's frame of reference. The
effect simply follows from the unsteady change in the propagation delay of
photons which distorts the transformation of time between the planet and the
observer.  The variation rate of the delay is changing most rapidly at
the middle of the transit when the full Newtonian acceleration vector
points straight along the line-of-sight. In the following sections, we
derive the magnitude of the resulting lightcurve asymmetry (\S 2) and
discuss its implications (\S 3).

\section{Apparent Transverse Deceleration Due to Propagation Delay of Photons}

The observed arrival time of photons $t_{\rm obs}$ is given by the time
they left the planet $t$ plus the propagation delay over the distance of
the planet $D$,
\begin{equation}
t_{\rm obs}=t+ D/c .
\label{eq:Doppler}
\end{equation}
The $t$-derivative of equation~(\ref{eq:Doppler}) gives
\begin{equation}
dt_{\rm obs}=dt \left(1 + {\vpar\over c}\right) ,
\label{eq:deriv}
\end{equation}
where $\vpar (t)=dD/dt$ is the velocity of the planet along the
line-of-sight. Since $\vpar$ changes from negative to positive during the
transit of a planet in a circular orbit, it is now obvious that ingress
would advance at a faster rate (i.e. a shorter $dt_{\rm obs}$ per $dt$
interval) than egress. Figure 1 illustrates schematically this generic
behaviour.  Note that equation (\ref{eq:deriv}) is accurate to all orders
in $\vert {\bf v}/c \vert$ but we will keep only first order terms in
subsequent derivations.  Even though the planet is not emitting the
observed photons, the timing of its occultation is dictated by the instant
at which stellar photons graze its outer surface; this timing is distorted
by the same propagation delay effect that would exist if the grazing
photons were emitted by the planet itself rather than the star since the
propagation history of the photons before they graze the planet's boundary
is irrelevant.  The transverse spatial coordinates of the planet orbit,
${\bf x}_\perp$, are the same in the observer and planet reference frames.
The apparent transverse velocity of the planet is therefore different from
its actual transverse velocity, $\vperp = ({d {\bf x}_\perp/ dt})$
\citep{RL},
\begin{equation}
{\bf v}_{\perp, {\rm obs}}\equiv {d {\bf x}_\perp\over dt_{\rm obs}}=
\left({dt\over dt_{\rm obs}}\right){d {\bf x}_\perp\over dt}\approx 
\left(1 - {\vpar \over c}\right) \vperp  ,
\label{eq:velocity}
\end{equation}
where the transverse position vector ${\bf x}_\perp$ corresponds to angular
coordinates on the sky times the distance to the planetary
system. Throughout our discussion, the terms parallel ($\parallel$) or
transverse ($\perp$) are relative to the line-of-sight axis that starts at
the observer and goes through the center of the stellar image. We focus on
the time when the occulting planet crosses this spatial point of symmetry.

\begin{figure}
\includegraphics[angle=-90,scale=0.37]{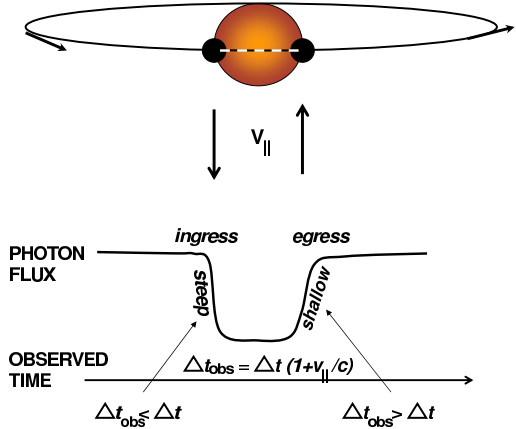}
\caption{Schematic illustration of the propagation delay effect.  The
transiting planet is moving towards the observer ($\vpar <0$) during
ingress and away from the observer ($\vpar >0$) during egress. As a result,
the observed time interval of partial eclipse is shorter at ingress
relative to egress (see Eq. \ref{eq:deriv}). This breaks the time reversal
symmetry of the observed lightcurve and introduces a fractional difference
of $\delta \vpar/c$ in the {\it temporal slope} of the lightcurve between
ingress and egress, where $\delta \vpar=\apar\tau$ is the net gain in the
planet's line-of-sight velocity over the transit duration $\tau$. For
close-in planets, the fractional asymmetry in the slopes is of order
$\delta\sim 10^{-4}$--$10^{-3}$ (see Eq. \ref{eq:min}).}
\label{fig1}
\end{figure}

Taking the $t_{\rm obs}$--derivative of both sides of
equation~(\ref{eq:velocity}) and keeping terms to leading-order in $\vert
{\bf v}/c\vert$, we get two new contributions to the difference between the
observed and Keplerian values of the transverse acceleration of the planet
across the sky \citep{Loeb},
\begin{equation}
{\bf {\dot v}}_{\perp, {\rm obs}}= \aperp - {2\vpar \over c} \aperp - {{\bf
v}_\perp\over c} \apar ,
\label{eq:acceleration}
\end{equation}
where ${\bf \dot{v}}_{\rm obs}\equiv ({d^2 {\bf x}/dt^2_{\rm obs}})$ is the
observed acceleration and ${\bf \dot{v}} \equiv ({d^2 {\bf x}/ dt^2})$ is
the actual Keplerian acceleration of the planet. The last term on the
right-hand-side of equation (\ref{eq:acceleration}) implies that the
apparent {\it transverse} acceleration of the planet gets a contribution
from its Keplerian acceleration {\it along} the line-of-sight, $\apar$.
This term is the source of our effect.

For simplicity, we assume that the orbital plane is viewed edge-on by the
observer; this geometry is exact for transits that cross the center of the
star (``central transits'') and is a very good approximation more generally
as long as the orbital radius is much larger than the radius of the star.
For a circular planetary orbit, Newtonian dynamics implies no transverse
acceleration of the planet at the spatial center of symmetry of the transit
where
\begin{equation}
\aperp=0~;~v_{\perp}^2={GM_\star\over a}~;~\apar={GM_\star \over
a^2}~;~\vpar=0.
\label{eq:sym}
\end{equation}
Here $M_\star$ is the stellar mass and $a$ is the planet's orbital radius.
However, equation~(\ref{eq:acceleration}) implies that at the same time
there would be an apparent transverse deceleration of the planet in the
observer's frame of reference
\begin{equation}
{\bf \dot{v}}_{\perp,{\rm obs}}= -{{\bf v}_\perp\over c} \apar =
- {\vperp\over \vert\vperp\vert}{(GM_\star)^{3/2}\over a^{5/2}c},
\label{eq:DopA}
\end{equation}
which is directed opposite to $\vperp$.  This persistent apparent
deceleration implies that the planet would cross ingress and egress at
different apparent rates.  Thus, even if the stellar image is perfectly
symmetric, the orbit is circular and the planet is perfectly spherical,
there is an inherent asymmetry in the transit lightcurve due to the nearly
steady value of $\apar=GM_\star/a^2$ during the transit.  Newtonian
dynamics alone predicts a non-vanishing $\aperp$ as soon as the planet
moves away from the transit center, but for a circular orbit this deviation
would maintain the time-reversal symmetry between ingress and egress.

The fractional asymmetry in the transit lightcurve is of order the
fractional change in $v_{\perp,{\rm obs}}$ over the transit duration, since
the rates by which ingress and egress proceed are proportional to
$v_{\perp,{\rm obs}}$. For a total transit duration $\tau$ and an orbital
period $T=2\pi a/v_\perp\gg \tau$, the slope of the initial drop and final
rise in the transit lightcurve will differ by a fractional amplitude
\begin{eqnarray}
\delta \equiv {\delta v_{\perp,\rm obs} \over v_{\perp,\rm obs}}& = & \vert
{{\bf \dot{v}}_{\perp,\rm obs} \tau\over {\bf v}_{\perp,\rm
obs}}\vert={\apar\tau\over c}= \left({v_\perp \over c}\right) \left({2\pi
\tau\over T}\right)
\label{eq:min}
\\ = 1.078 \times &10^{-4}& \left({M_\star\over 1.1 M_\odot}\right)
\left({a\over 7 \times 10^{11}~{\rm cm}}\right)^{-2}\left({\tau\over
3~{\rm hr}}\right) . \nonumber
\end{eqnarray}
For a ``hot Jupiter'' like \HD, the ingress phase ($\vpar <0$) would
proceed at a rate that is higher by fractional amplitude of $\sim 10^{-4}$
than the egress phase ($\vpar >0$). Planets that are closer in by a factor
of a few could produce an asymmetry of up to $\delta \sim 10^{-3}$;
OGLE-TR-56b for which $a=3.5\times 10^{11}~{\rm cm}$ provides $\delta
=3\times 10^{-4}$ [see \citet{Gau} for a compilation of all known transit
systems and the prospects for detecting others].

The observed time $t_{\rm obs}$ is a slightly distorted version of the time
axis $t$ along which the lightcurve is symmetric. We may write $t_{\rm
obs}=t(1+\epsilon)$, where $\epsilon(t)=(t/2\tau)\delta$ and we shifted
$t=0$ to be at the transit centroid.  The observed photon flux, $F(t_{\rm
obs})$, corresponds to the flux at an undistorted time $t\approx t_{\rm
obs}[1-\epsilon(t)]$, while $F(-t_{\rm obs})$ corresponds to $t\approx
-t_{\rm obs}[1-\epsilon(-t)]=-t_{\rm obs}[1+\epsilon(t)]$.  For small
deviations, we may expand the photon flux as a function of time to leading
order, $F(t+\Delta t)\approx F(t)+ (dF/dt)\vert_{t} \Delta t$.  Since
$\epsilon(t)=-\epsilon(-t)$, $F(t)=F(-t)$, and
$(dF/dt)\vert_t=-(dF/dt)\vert_{-t}$, the time-dependent photometric
asymmetry of the lightcurve is given by
\begin{eqnarray}
\Delta_F (t_{\rm obs})&\equiv& {F(t_{\rm obs})-F(-t_{\rm obs})\over F(t_{\rm
obs})} = \\
\label{eq:flux}
&=& -\left({d\ln F\over d\ln t_{\rm obs}}\right)_{t_{\rm obs}} \times
\left({t_{\rm obs}\over \tau}\right) \delta , \nonumber
\end{eqnarray}
where we have set $t_{\rm obs}=0$ at the transit centroid (for which
$dF/dt_{\rm obs}=0$) and kept terms to leading order in
$\vert\epsilon\vert\ll1$. Given a preliminary transit lightcurve, $F(t_{\rm
obs})$, and an approximate knowledge of the system parameters, it is
possible to predict the lightcurve asymmetry under the assumption of a
circular orbit. Figure 2 shows the expected $\Delta_F(t_{\rm obs})$ for \HD
based on the lightcurve data in \citet{Bro}.  For a transit depth of
$2\times 10^{-2}$ magnitudes produced by a close-in planet, the net
photometric asymmetry would typically amount to a few percent of $\delta$
or equivalently to a peak value of $\vert \Delta_F\vert$ in the range of
(0.1--1)$\times 10^{-5}$.  The photometric sensitivity of $\sim 10^{-4}$
magnitudes per data point achieved with {\it HST} for \HD \citep{Bro}
provides a net sensitivity to a time-reversal asymmetry of
$\vert\Delta_F\vert\sim 10^{-5}(N/10^2)^{-1/2}$, where $N\sim 10^2$ is the
total number of independent data points during ingress and egress. Our
effect appears to be on the borderline of being detectable with existing
techniques. A statistically significant ($\ga 3\sigma$) detection of the
asymmetry in equation~(\ref{eq:min}) for \HD would require an additional
order of magnitude improvement in sensitivity, which is achievable with
future space missions such as {\it
Kepler}\footnote{http://www.kepler.arc.nasa.gov/summary.html} or {\it
COROT}
\footnote{http://serweb.oamp.fr/projets/corot/} and only marginally with
the existing instrument {\it MOST}
\footnote{http://www.astro.ubc.ca/MOST/}. A simple way to reach the
required sensitivity would be to maintain the existing photometric
precision per data point and increase the number of data points from $\sim
10^2$ up to $\sim 10^4$ (e.g. by increasing the number of \HD transits from
the 4 observed by {\it HST} to $\sim 400$) over a period of several years.

\begin{figure}
\plotone{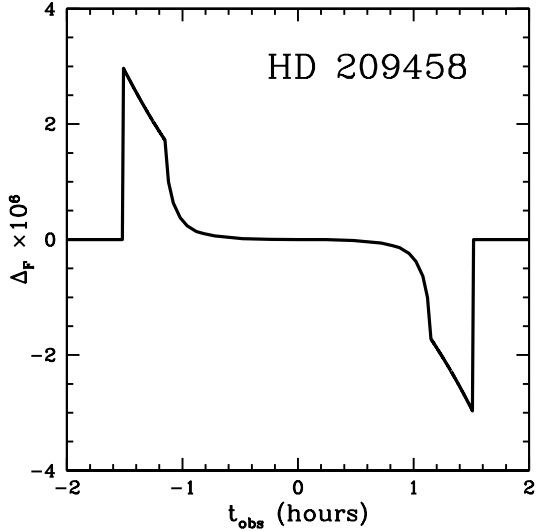}
\caption{Predicted fractional deviation from time-reversal symmetry,
$\Delta_F(t_{\rm obs})$, for the observed transit lightcurve of \HD
\citep{Bro}, based on Eqs. (\ref{eq:min}) and (8).  The vertical axis is
proportional to $(M_\star/1.1M_\odot)(a/7\times 10^{11}{\rm cm})^{-2}$. The
four breaks in the curve result from the four points of contact between the
projected planet boundary and the limb of the star.}
\label{fig2}
\end{figure}

The orbital period scales as $\propto (M_\star/a^3)^{-1/2}$ while the
fractional asymmetry in equation~(\ref{eq:min}) scales as $\propto
M_\star/a^2$.  Thus, detection of the lightcurve asymmetry would allow to
determine $M_\star$ and $a$ separately. Additional data on the reflex
motion of the star would then provide the planet mass, $M_{\rm P}$, based
on dynamical information alone. Currently, the stellar mass, $M_\star$, is
inferred indirectly based on theoretical fitting of the stellar spectrum
which is subject to modeling uncertainties at the level of $\ga 10\%$ [see
Fig. 3 in \citet{Alo}].

If the planetary orbit is nearly circular but has a finite eccentricity,
$e\ll 1$, then it is easy to show that at the transit center
\begin{equation}
\dot{v}_\perp= - e\sin \Psi {GM_\star\over a^2},
\label{eq:ecc}
\end{equation}
to leading order in $e$, where $\Psi$ is the angle between the shortest
radius vector of the orbital ellipse (along the apsidal line) and the line
to the observer. As long as $\vert e\sin \Psi\vert < \vert
v_{\perp}/c\vert\la 10^{-3}$, the eccentricity effect is smaller than the
propagation delay effect in equation~(\ref{eq:DopA}).  In the absence of
perturbers, any initial orbital eccentricity is expected to decay
exponentially through tidal dissipation on an e--folding timescale
\citep{Gold}
\begin{eqnarray}
t_{\rm circ}& = & {e\over \dot{e}}=\left({4T Q_{\rm P}\over 63\times
2\pi}\right) \left({M_{\rm P}\over M_\star}\right)\left({a\over R_{\rm
P}}\right)^5=
\label{eq:dis}
\\  =  1.4&\times & 10^{-2} \left({Q_{\rm P}\over 10^5}\right)\left({M_{\rm
P}\over 6\times 10^{-4}M_\star}\right)\left({a\over 75 R_{\rm
P}}\right)^5~{\rm Gyr}, \nonumber
\end{eqnarray}
where the quality parameter $Q_{\rm P}$ ($\sim 10^5$ for Jupiter) is
inversely proportional to the dissipation rate in the planet's interior
\citep{Ioa,Ogi}, and $R_{\rm P}$ is the planet's radius.  Aside from the
unknown $Q_{\rm P}$, the parameter values in equations (\ref{eq:min}) and
(\ref{eq:dis}) were chosen to match \HD. Since the orbital circularization
timescale is shorter by up to two orders of magnitude than the age of
planetary systems such as HD 209458 [$\sim 5$ Gyr, see \citet{Maz} and
\citet{Cod}], the eccentricity-driven asymmetry is expected to diminish for
a close-in planet unless a third body pumps its orbital eccentricity.
Incidentally, such a perturber was hypothesized as the driver of the
inflated radius of \HD \citep{Bod} but its existence has not been
demonstrated yet.

For close-in planets such as \HD, our effect is larger than a different
source of lightcurve asymmetry that was already discussed in the
literature, namely the planet's obliquity \citep{Hui2}.  This is because
close-in planets are expected to rotate slowly and possess a small
projected obliquity as a result of strong tidal locking with their orbital
revolution. Their rotation axis is expected to be normal to their orbital
plane which is viewed nearly edge-on [in the case of \HD , the orbital
inclination angle is $i=86.^\circ1\pm 1.^\circ6$ \citep{Maz}].  Another
source of asymmetry is rotation of the star. Typical projected rotation
speeds of $v_{\rm rot}\sin i\sim 4~{\rm km~s^{-1}}$ \citep{Maz,Que} would
produce a Doppler offset \citep{Loe2} between the flux emitted by the
approaching and receding sides of the star of order $\sim (v_{\rm
rot}/c)=1.3\times 10^{-5}$ [the photometric analog of the {\it
Rossiter-McLaughlin effect}; see \citet{Que,Oht}], or an occultation
contrast of $\sim 10^{-7}$ magnitudes which is much smaller than the effect
considered here.  If the stellar rotation axis is significantly misaligned
with the normal of the orbital plane and the transit is not central, then
the rotation-induced oblateness of the stellar image could generate a
fractional ($\delta$--equivalent) asymmetry of less than ${1\over 2}(v_{\rm
rot}^2R_\star /GM_\star) \sim 5\times 10^{-5}$, which is again well below
our effect for \HD where the misalignment angle must be small \citep{Que}.

The propagation delay relation between $t_{\rm obs}$ and $t$ in equation
(\ref{eq:Doppler}) can be easily incorporated into a computer program that
searches for the best-fit Keplerian orbit under the constraints of a given
data set. The {\it delay--corrected} Keplerian fit would provide new
dynamical constraints on the planetary system. Such a fit would involve the
same number of free parameters as the standard Keplerian fit.

\section{Discussion}

We have shown that the propagation delay of light introduces an {\it
apparent} transverse deceleration of a planet on the sky during its transit
across the face of its parent star (Eq. \ref{eq:DopA}).  This persistent
deceleration breaks the time-reversal symmetry of the transit lightcurve
for a spherical planet in a circular orbit around a spherical
star. Throughout the transit, the planet velocity along the line-of-sight,
$\vpar$, is changing at a nearly steady rate, $GM_\star/a^2$. This produces
a steady change in the transformation of time to the observer's frame (see
Eq. \ref{eq:deriv}).  The net change in $\vpar$ between ingress and egress,
$\delta \vpar= \apar\tau$, introduces a fractional difference of magnitude
$\delta=\delta \vpar/c$ in the slopes of their lightcurves. It is possible
to search for this difference in slopes by folding the lightcurve over its
centroid.  For close-in planets, ingress should typically proceed at a rate
that is faster by a fractional amplitude $\delta \sim 10^{-4}$--$10^{-3}$
than egress (see Eq. \ref{eq:min}). This level of asymmetry will be
detectable with forthcoming space telescopes such as {\it COROT} or {\it
Kepler} \citep{Bor}, which are scheduled for launch within 2-3 years.
Because this asymmetry has a unique scaling with stellar mass and orbital
radius ($\propto M_\star/a^{2}$), its detection together with the reflex
motion of the star will allow to determine the star and planet masses as
well as the orbital radius using purely dynamical data. This method evades
the theoretical uncertainties inherent in the existing approach for
determining stellar masses based on modeling of spectroscopic data
[e.g. \citet{Cod}].

Orbital eccentricity could induce a stronger asymmetry in the lightcurve
but is expected to decay exponentially to negligible levels through tidal
dissipation for close-in planets like \HD, unless it is being pumped by the
gravitational perturbation of another planet.

Similarly to other transit timing residuals \citep{Mir,Hol,Ago}, our effect
would be contaminated by noise from inhomogeneities on the face of the star
and would compete against other small effects involving asymmetries from
the oblateness of the planet \citep{Hui,Hui2} or the rotation of the star.
Additional special-relativistic or general-relativistic effects are of the
order of $\sim (v/c)^2$ or $\sim \phi/c^2$, or smaller, where $\phi$ is the
gravitational potential produced by the star ($\phi \sim v^2$); these
corrections are orders of magnitude smaller than the propagation delay
effect discussed here.

Finally, we note that a change of the opposite sign in $\vpar$ occurs when
the planet goes behind the star.  In this case, ingress would be slower
than egress.  There is no net change in the orbital period over a full
closed orbit.  However when the planet enters its own (secondary) eclipse
by the star, the photons it emits will be delayed relative to primary
eclipse (the transit) by the difference in emission times plus the light
travel time across the orbital diameter, as implied by
equation~(\ref{eq:Doppler}).  For a circular orbit, the time interval
between the centroids of the primary and secondary eclipses would be longer
by $\delta T_{1/2}=2a/c$ than half of the full orbital period.  
Future detection of a planet's infrared emission would then allow to
determine the orbital radius from the light propagation delay of $\delta
T_{1/2}=46.7 (a/7\times 10^{11}~{\rm cm})$ seconds. The full orbital period
$T$ would then yield the stellar mass through $M_\star=a^3/[G(T/2\pi)^2]$.
An eccentricity could change the lightcurve history of the illuminated
planet and in particular make the time from primary eclipse to secondary
eclipse different from the time back to primary eclipse.

\acknowledgments

I thank David Charbonneau, Scott Gaudi, Matt Holman, Dimitar Sasselov,
George Rybicki, and Josh Winn for helpful discussions. This work was
supported in part by NASA grant NAG 5-13292 and NSF grants AST-0071019 and
AST-0204514.


\begin{thebibliography}{}

\bibitem[Agol et al.(2004)]{Ago} Agol, E., Steffen, J., Sari, R., \&
Clarkson, W. 2004, MNRAS, submitted; astro-ph/0412032

\bibitem[Alonso et al.(2004)]{Alo} Alonso, R., et al. 2004, ApJ, 613, L153

\bibitem[Bodenheimer et al.(2003)]{Bod} Bodenheimer, P., 
Laughlin, G., \& Lin, D.~N.~C.\ 2003, \apj, 592, 555 

\bibitem[Borucki et al.(2003)]{Bor} Borucki, W. J., et al. 2003, in Future
EUV/UV and Visible Space Astrophysics Missions and Instrumentation,
Ed. J. C, Blades \& O. H. W. Siegmund, Proc. of SPIE, Volume 4854, 
pp. 129-140

\bibitem[Bouchy et al.(2004)]{Bou} Bouchy, F., Pont, F., Santos, N.~C.,
Melo, C., Mayor, M., Queloz, D., \& Udry, S.\ 2004, \aap, 421, L13

\bibitem[Brown et al.(2001)]{Bro} Brown, T.~M., Charbonneau, D., Gilliland,
R.~L., Noyes, R.~W., \& Burrows, A.\ 2001, \apj, 552, 699

\bibitem[Charbonneau et al.(2000)]{Cha} Charbonneau, D., Brown, T. M.,
Latham, D. W., \& Mayor, M.  2000, ApJ, 529, L45

\bibitem[Cody \& Sasselov(2002)]{Cod} Cody, A.~M., \&
Sasselov, D.~D.\ 2002, \apj, 569, 451


\bibitem[Gaudi et al.(2005)]{Gau} Gaudi, S. , Seager, S., \&
Mallen-Ornelas, G. 2005, ApJ, in press; astro-ph/0409443

\bibitem[Goldreich \& Soter(1966)]{Gold} Goldreich, P., \& Soter, S. 1966,
Icarus, 5, 375

\bibitem[Holman \& Murray(2004)]{Hol}
Holman, M., \& Murray, N. W. 2004, Science, submitted; astro-ph/0412028 

\bibitem[Hui \& Seager(2002a)]{Hui} Hui, L., \& Seager, S.\ 
2002a, \apj, 572, 540 

\bibitem[Hui \& Seager(2002b)]{Hui2} -------------------------. 
2002b, \apj, 574, 1004 

\bibitem[Ioannou \& Lindzen(1993)]{Ioa} Ioannou, P.~J., \& 
Lindzen, R.~S.\ 1993, \apj, 406, 266 

\bibitem[Konacki et al.(2004a)]{Kon} Konacki, M. et al. 2004a,
ApJ, 609, L37

\bibitem[Konacki et al.(2004b)]{Kon2} Konacki, M.. Torres, G., Sasselov, D.,
\& Jha, S. 2004b, astro-ph/0412400

\bibitem[Loeb(2003)]{Loeb} Loeb, A. 2003, astro-ph/0309716

\bibitem[Loeb \& Gaudi(2003)]{Loe2} Loeb, A., \& Gaudi, 
B.~S.\ 2003, \apjl, 588, L117 

\bibitem[Mazeh et al.(2000)]{Maz} Mazeh, T., et al. 2000, ApJ, 532, L55

\bibitem[Miralda-Escud{\' e}(2002)]{Mir}
Miralda-Escud{\'e}, J.\ 2002, \apj, 564, 1019

\bibitem[Moutou et al.(2004)]{Mou}
Moutou, C., Pont, F., Bouchy, F., \& Mayor, M. 2004, A \& A, in press;
astro-ph/0407635

\bibitem[Ogilvie \& Lin(2004)]{Ogi} Ogilvie, G.~I., \& Lin, 
D.~N.~C.\ 2004, \apj, 610, 477

\bibitem[Ohta et al.(2004)]{Oht}
Ohta, Y., Taruya, A., \& Suto, Y. 2004, astro-ph/0410499

\bibitem[Pont et al.(2004)]{Pon} Pont, F., Bouchy, F., Queloz, D., Santos,
N. C., Melo, C., Mayor, M., \& Udry, S. 2004, A \& A, 426, L15

\bibitem[Queloz et al.(2000)]{Que} Queloz, D., Eggenberger, 
A., Mayor, M., Perrier, C., Beuzit, J.~L., Naef, D., Sivan, J.~P., \& Udry, 
S.\ 2000, \aap, 359, L13 


\bibitem[Rybicki \& Lightman(1979)]{RL} Rybicki, G. B. \& Lightman,
A. P. Radiative Processes in Astrophysics, (New York: Wiley, 1979), see
problem 4.7a on p. 151

\bibitem[Torres et al.(2004)]{Tor} Torres. G., Konacki, M., Sasselov, D., 
Jha, S. 2004, ApJ, 609, 1071


\end{thebibliography}
\end{document}